\documentclass[10pt,preprint]{aastex} 
\usepackage{lscape}

\newcommand{\ovi}{O$\;${\small\rm VI}\relax}
\newcommand{\siiv}{Si$\;${\small\rm IV}\relax}
\newcommand{\civ}{C$\;${\small\rm IV}\relax}
\newcommand{\feii}{Fe$\;${\small\rm II}\relax}

\newcommand{\htwo}{H$_2$}

\newcommand{\HI}{H$\;${\small\rm I}\relax}

\newcommand{\kms}{km~s$^{-1}$\relax}

\newcommand{\percc}{cm$^{-3}$\relax}
\newcommand{\column}{cm$^{-2}$}
\newcommand{\nav}{$N_a(v)$}

\newcommand{\dg}{$^\circ \,$}
\newcommand{\sk}{Sk--}

\newcommand{\err}[2]{^{+#1}_{-#2}}

\newcommand{\fuse}{{\em FUSE}}

\newcommand{\hst}{{\em HST}}
\newcommand{\iue}{{\em IUE}}

\begin{document}

\slugcomment{\bf To Appear in {\em The Astrophysical Journal}}

\title{{\em FUSE} Observations of Degree-Scale Variations in 
	Galactic Halo \ovi}

\author{J. Christopher Howk\altaffilmark{1}, 
	Blair D. Savage\altaffilmark{2}, 
	Kenneth R. Sembach\altaffilmark{1,3},
	\& Charles G. Hoopes\altaffilmark{1}}

\altaffiltext{1}{Department of Physics and Astronomy,
The Johns Hopkins University, Baltimore, MD, 21218; howk@pha.jhu.edu,
choopes@pha.jhu.edu}

\altaffiltext{2}{Astronomy Department, University of Wisconsin-Madison,
Madison, WI, 53711; savage@astro.wisc.edu}

\altaffiltext{3}{Current address: Space Telescope Science Institute, 
3700 San Martin Dr., Baltimore, MD 21218; sembach@stsci.edu}

%%%%%%%%%%%%%%%%%%%%%%%%%%%%%%%%%%%%%%%%%%%%%%%%%%%%%%%%%%%%%%%%%%%%%%%

\begin{abstract}

We report {\em Far Ultraviolet Spectroscopic Explorer} (\fuse)
observations of interstellar \ovi\ absorption in the halo of the Milky
Way towards 12 early-type stars in the Large Magellanic Cloud (LMC)
and 11 in the Small Magellanic Cloud (SMC).  The mean column densities
of \ovi\ associated with the Galactic halo towards the LMC and SMC are
$\log \langle N(\mbox{\ovi}) \rangle = 14.52\err{0.10}{0.14}$ and
$14.13\err{0.14}{0.20}$, respectively, where the uncertainties
represent the standard deviations of the individual measurements about
the means.  Significant variations in the \ovi\ column densities are
observed over all of the angular scales probed by our observations:
$0\fdg5 - 5\fdg0$ towards the LMC and $0\fdg05 - 3\fdg3$ towards the
SMC.  The maximum factors by which the \ovi\ varies between sight
lines towards the LMC and SMC are $\sim2.8$ and $\sim4.2$,
respectively.  Though low-, intermediate-, and high-velocity clouds
are present along most of the sight lines towards the LMC, the
variations in \ovi\ column densities are found in all of these \ovi\
absorbing structures.  The column density variations trace the
structure of the hot ionized medium in the Galactic halo on scales of
$\ll 80$ to 800 pc towards the LMC and $\ll 6$ to 400 pc towards the
SMC (assuming the absorption arises within the first 5 kpc above the
Galactic plane).

\end{abstract}


%%%%%%%%%%%%%%%%%%%%%%%%%%%%%%%%%%%%%%%%%%%%%%%%%%%%%%%%%%%%%%%%%%%%%%%

%LMC: (l,b) = (280.5, -32.9); vr ~ +264; d ~ 50 kpc

%SMC: (l,b) = (302.8, -44.3); vr ~ +165; d ~ 65 kpc

\section{Introduction}
\label{sec:intro}

\defcitealias{p1}{Paper~I}
\defcitealias{p2}{Paper~II}

Observations of absorption from \ovi\ in the interstellar medium (ISM)
towards early-type stars and extragalactic objects provide a useful
means of studying the hot gaseous medium of galaxies.  Because 113.9
eV are required to ionize \ion{O}{5} to \ovi, photoionization is
unlikely produce the observed quantities of \ovi\ in galactic
environments \cite[see][]{savage00,sembach00}.  This fact, coupled
with the high intrinsic abundance of O and the strength of the \ovi\
absorption doublet at 1031.926 and 1037.617 \AA, makes this ion a
sensitive probe of the conditions of the hot ISM in the Milky Way and
other nearby galaxies.

%The \copernicus\ observatory was the first to detect interstellar
%\ovi\ absorption \citep{jenkinsmeloy74,york74,jenkins78a,jenkins78b}.  
%With the exception of the short-lived ORFEUS-SPAS platform \cite[see
%observational work by][]{hurwitzbowyer96,hurwitz95,widmann98,hurwitz98,
%sembachsavagehurwitz99}, high resolution observations of the
%far-ultraviolet (FUV) \ovi\ doublet have not been possible since the
%end of the \copernicus\ mission 20 years ago.  With the launch of the
%{\em Far Ultraviolet Spectroscopic Explorer} (\fuse) in June 1999 the
%FUV spectrum has been reopened to high-resolution ($R\sim15,000 -
%20,000$) spectroscopy in the wavelength range 905--1187 \AA.
%\fuse\ can easily observe stars in the Galactic halo, stars in the 
%Magellanic Clouds, and distant extragalactic sources with sufficient
%signal to noise to produce reliable \ovi\ column density and line
%profile measurements.  Early reports of \fuse\ observations include
%studies of interstellar \ovi\ absorption in the Galactic halo
%\citep{savage00,oegerle00}, in high-velocity clouds
%\citep{sembach00,murphy00}, and in the Galaxy and the
%Large Magellanic Cloud towards the star \sk67\dg05 \citep{friedman00}.

The {\em Far Ultraviolet Spectroscopic Explorer} (\fuse), launched in
June 1999, is producing a large database of high-resolution
($R\sim15,000 - 20,000$) spectra covering the far-ultraviolet (FUV)
wavelength range 905--1187 \AA, which includes the important \ovi\
doublet.  \fuse\ can easily observe stars in the Galactic halo, stars
in the Magellanic Clouds, and distant extragalactic sources with
sufficient signal to noise to produce reliable \ovi\ column density
and line profile measurements.  Early reports of \fuse\ observations
include studies of interstellar \ovi\ absorption in the Galactic halo
\citep{savage00,oegerle00}, in high-velocity clouds
\citep{sembach00,murphy00}, and in the Galaxy and the
Large Magellanic Cloud towards the star \sk67\dg05 \citep{friedman00}.

In this paper we discuss \fuse\ observations of interstellar \ovi\ in
the halo of the Milky Way as observed towards 12 early-type stars in
the Large Magellanic Cloud [LMC; $(l,b) = (280.5, -32.9)$; $d=50$ kpc]
and 11 early-type stars in the Small Magellanic Cloud [SMC; $(l,b) =
(302.8, -44.3)$; $d=65$ kpc].  These observations have previously been
used to study the \ovi\ content of these galaxies \citep{p1,p2}.  Our
emphasis in this paper is on the degree of variability in the \ovi\
column densities over the small angular scales ($0\fdg05 \le \Delta
\theta \le 5\fdg1$) probed by these Magellanic Cloud observations.  

The remainder of this paper is laid out as follows.  In
\S\ref{sec:sightlines} we discuss the \fuse\ observations and the 
properties of the sight lines towards each of the Magellanic Clouds.
We present the small-scale \ovi\ measurements in
\S\ref{sec:variations}, and we briefly discuss the kinematics of the
\ovi\ profiles along these directions in \S\ref{sec:kinematics}.  We
discuss the implications of these measurements for the the
highly-ionized Galactic halo in \S\ref{sec:discussion} and summarize
our results in \S\ref{sec:summary}.

\section{\fuse\ Observations of the Magellanic Cloud Sight Lines}
\label{sec:sightlines}

\subsection{The LMC Sight Line}
\label{subsec:lmcsightline}

The first high-resolution UV absorption line measurements along sight
lines towards extragalactic objects were were {\em International
Ultraviolet Explorer} (\iue) observations of early-type stars in the
LMC \citep{savagedeboer79}.  This high-latitude sight line
($b\sim-33^\circ$) probes the thin interstellar disk of the Milky Way
only along the first few hundred parsecs ($z\sim200$ pc at a distance
of $d\sim400$ pc from the Sun). The first $\sim100$ pc in this
direction are contained within the boundaries of the local bubble,
where there is relatively little material
\citep{sfeir99}.  Most of the LMC sight line probes the ``thick disk'' or 
``halo'' of the Milky Way, and many of the interstellar clouds along
this sight line show abundances reminiscent of other Galactic halo
clouds \citep{welty99,friedman00}.  Gas participating in Galactic
rotation \citep{clemens85} within the first $\sim5$ kpc of the Sun ($z
\la 2.5$ kpc) in this direction will be at low negative velocities
($-5 \la v_{\rm LSR} \la 0$ \kms).  More distant gas will be found at
positive velocities, reaching $v_{\rm LSR} \sim +100$ \kms\ at 25 kpc
($z\sim12.5$ kpc).

Several clouds or cloud complexes have been identified along the sight
line to the LMC, mostly in tracers of neutral gas.  \citet*{mcgee83}
summarize \HI\ 21-cm observations of the cloud complexes in the
general direction of the LMC.  They find several components towards
the LMC.  The main emission complexes are seen at $\langle v_{\rm LSR}
\rangle = -32.8, +0.2, +65.5, \ {\rm and} \ +131.7$ \kms, where the
velocities are a spatial average over the face of the LMC, with
spatially-averaged \HI\ column densities of $7.1\times10^{18}$,
$4.3\times10^{20}$, $4.3\times10^{18}$, and $6.0\times10^{18}$ atoms
\column.  The component nearest the LSR can be associated with local
gas in the Galactic disk.  The $\langle v_{\rm LSR} \rangle \sim -33$
and $+65$ \kms\ components are referred to as intermediate velocity
clouds (IVCs), while the $\sim+130$ \kms\ cloud is cataloged as a high
velocity cloud (HVC).  These components can also be seen in metal
absorption lines \citep*[e.g.][]{ferlet85, savagedeboer81}, including
moderately- and highly-ionized species \citep{welty99, wakker98}, and
\citet{danforth02} have discussed their distribution in
\fuse\ observations of metal absorption lines.  Each of these
aforementioned ``components'' is likely made up of several individual
clouds that are unresolved in many datasets \citep{welty99}, and the
IVC and HVC complexes probably reside in the Galactic halo
\citep{welty99}.

The highly-ionized gas along the sight line to the LMC at velocities
below those associated with the LMC itself likely traces material in
the vertically-extended Galactic halo.  \iue\ observations of \civ\
and \siiv\ along LMC sight lines have been presented by
\citet{sembachsavage92}, \citet{savage89}, \citet{savagedeboer81}, 
and others, while \citet{bomans96} and \citet{wakker98} have presented
\hst\ observations of Galactic \civ\ along LMC sight lines.  We also
note that \citet{widmann98} have derived \ovi\ column densities for
the Galactic halo seen towards the LMC using ORFEUS data.  However,
high-quality measurements of the highly-ionized Galactic material
towards the LMC have been sparse.

In this work we discuss the Galactic \ovi\ absorption as observed with
\fuse\ towards 12 early-type stars in the LMC.  The observations and 
analysis of the data used in this work are described in detail by
\citet[hereafter \citetalias{p1}]{p1} who make use of the observations
to study the hot gas content of the LMC.  All of the \fuse\
observations used for this work have a resolution of $\la20$ \kms\
(FWHM).  The LMC stars used as background continuum sources in
\citetalias{p1} are all either WR stars or O stars with spectral
types earlier than O7 and were chosen to have well-defined continua in
the region of interstellar \ovi\ $\lambda1031.926$ \AA\ absorption.
The (6-0) P(3) and (6-0) R(4) transitions of molecular hydrogen can
overlap Galactic \ovi\ absorption towards the LMC given the rest
wavelengths of these lines and the velocity structure of these sight
lines.  Models for molecular hydrogen transitions that overlap the
Galactic \ovi\ absorption have been divided out in the processing of
the spectra.  The exposure times for the LMC stars ranged from 3.6 to
33 ksec per star, yielding signal-to-noise ratios in the range
$\sim12$ to 25 per resolution element.  The continuum placement,
molecular hydrogen correction, and other details are described in
\citetalias{p1}.  

\begin{figure}
\epsscale{0.85}
\plotone{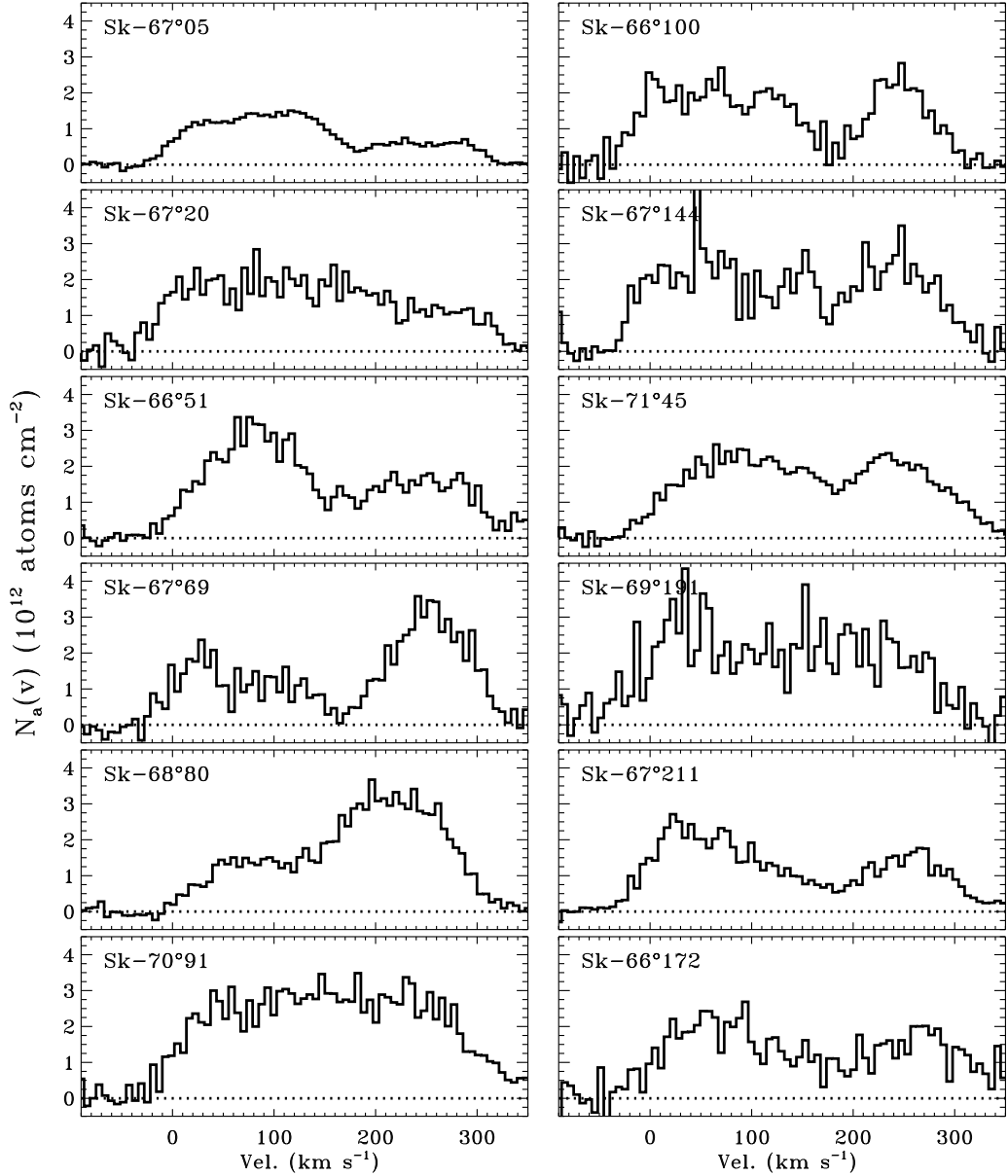}
\caption{Apparent column density, $N_a(v)$, versus velocity profiles for the
sight lines towards the 12 LMC probes discussed in \citetalias{p1}.
The LMC \protect\ion{O}{6} is at velocities $v \ga 175$
\protect\kms.  Material at velocities lower than this is associated 
with the Milky Way and the intermediate- and high-velocity clouds at
$v\sim+65$ and $+125$ km s$^{-1}$, respectively.
\label{fig:lmcnav}}
\end{figure}

Table \ref{tab:lmcovi} gives the Milky Way \ovi\ column densities
derived by integrating the apparent column density, \nav, profiles
\citep[see][]{savagesembach91} of the 1031.926 \AA\ transition.
The \nav\ profiles themselves are displayed in Figure
\ref{fig:lmcnav}.  Table \ref{tab:lmcovi} gives the \ovi\ column
densities integrated over two ranges (all velocities are approximately
LSR; see \citetalias{p1}): the first over the velocity range $|v| \le
50$ \kms, and the second over the entire velocity range of the Milky
Way (the full range is indicated in the last column of Table
\ref{tab:lmcovi}).  As discussed in \citetalias{p1} the separation
between the Milky Way and LMC absorption is not always clear.  In
these cases we adopt an average separation point of $+175$
\kms. The quoted column density errors for the columns integrated over
the entire range include the effects of changing the upper velocity
limit by $\pm20$ \kms, which is sufficient to cover the range of
separations observed in cases where there is a clear break between the
Milky Way and LMC absorption.

The full range of Milky Way velocities likely includes \ovi\
associated with the IVCs and HVCs present along the LMC sight lines.
Examining the strongest transitions of \feii\ and \ion{O}{1} in the
\fuse\ bandpass (at 1144.938 \AA\ and 1039.230 \AA, respectively), we
find that 11 of the 12 sight lines studied here exhibit IVC
absorption, while at least 10 of the 12 contain HVC
absorption.\footnote{\citet{danforth02} present the absorption line
profiles of several species towards almost 100 stars in the Magellanic
Clouds, including the strong \feii\ 1144.938
\AA\ and \ion{O}{1} 1039.230 \AA\ transitions.  We direct the reader
to that work for better views of the sight line velocity structure
along the lines of sight studied in this work.}  None of the sight
lines is simultaneously free of absorption from IVC and HVC material
(i.e., absorption from one or the other is seen along all of the sight
lines).  The column densities derived by integrating over $|v| \le 50$
\kms\ should exclude much of the IVC and HVC absorption.

\subsection{The SMC Sight Line}
\label{subsec:smcsightline}

The higher-latitude ($b\sim-44^\circ$) sight line to the SMC is more
heavily biased towards low-density Galactic halo gas than that towards
the LMC.  Towards the SMC, gas within $\sim10$ kpc ($z\la7$ kpc) that
is participating in Galactic rotation \citep{clemens85} would be found
at negative velocities, and gas within $\sim20$ kpc of the Sun
($z\la14$ kpc) should be found at velocities $|v_{\rm LSR}| \la 40$
\kms.

Similar to the case for the LMC, the sight line towards the SMC
contains a number of low-ionization Galactic gas clouds or complexes
spread over $\sim65$ \kms\ \citep{welty97,mallouris01}.  Material
associated with the disk and low halo of the Galaxy is seen over the
range $-25 \la v_{\rm LSR} \la +40$ \kms.  High-velocity Galactic
material, such as that seen towards the LMC, would be blended with SMC
material for velocities $v_{\rm LSR} \ga +60$ \kms.

Observations of the highly-ionized material in the direction of the
SMC with \iue\ are discussed by \citet{fitzsavage83} and
\citet{fitz84}.  More recently, \citet{koenigsberger01} and 
\citet{hoopes01} have reported on \hst\ and \fuse\ observations
of the sight line towards HD~5980 in the SMC.  These papers include
comments on the highly-ionized species \civ, \siiv, and \ovi\ in the
Galactic halo.  

In this work we make use of \fuse\ observations of Galactic \ovi\
towards 11 early-type stars in the SMC.  The observations and analysis
of the data are described in \citet[hereafter \citetalias{p2}]{p2},
who use these observations to study the hot gas content of the SMC.
The principal properties of the data are the same as those described
above for the LMC observations.

\begin{figure}
\epsscale{0.8}
\plotone{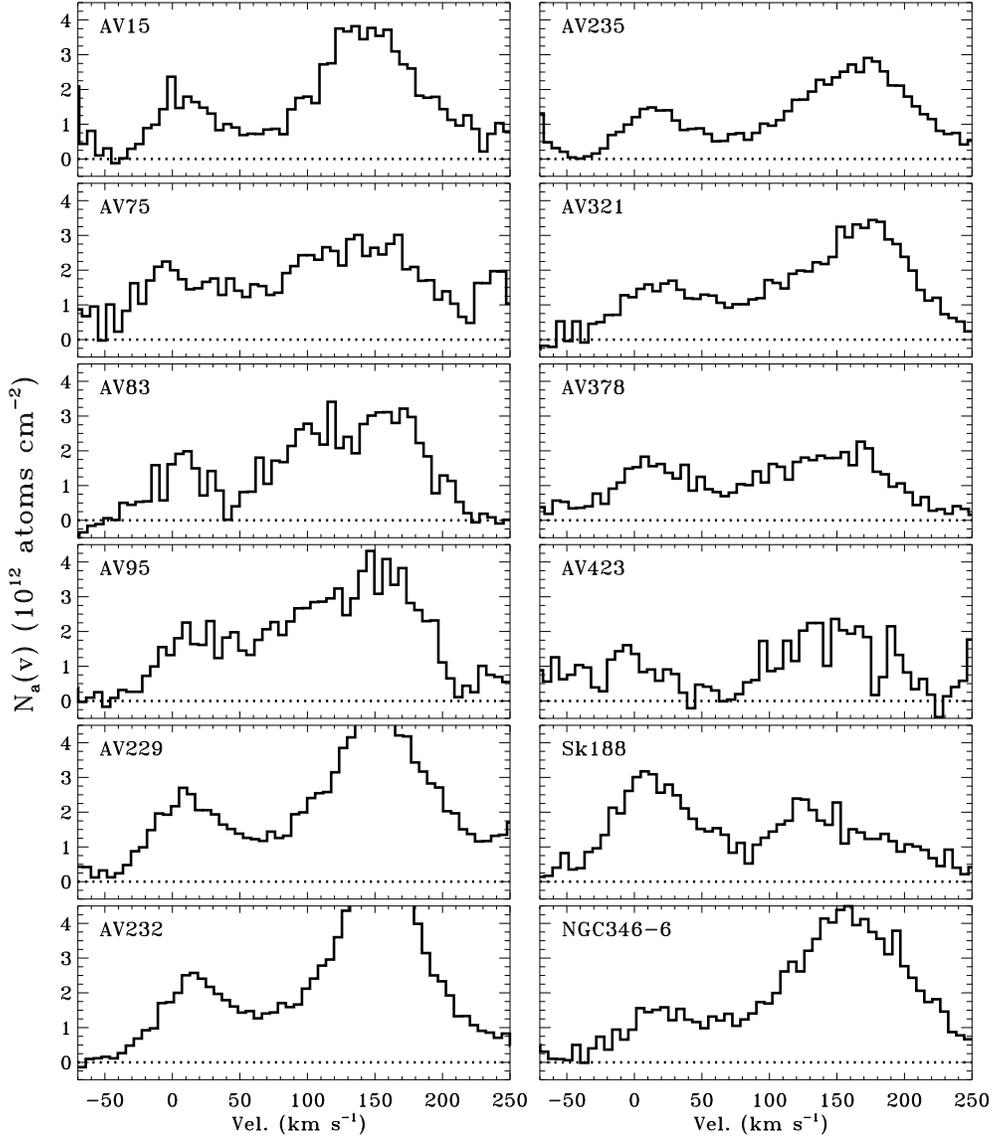}
\caption{Apparent column density, $N_a(v)$, versus velocity profiles for the
sight lines towards 12 of the SMC probes discussed in \citetalias{p2}.
The \protect\ion{O}{6} associated with the SMC is at velocities $v \ga
+60$ km s$^{-1}$ (see \citetalias{p2} for precise integration limits).
Material at velocities lower than this is associated with the Milky
Way.  The observations of NGC~346-6 include several other early-type
stars in the \protect\fuse\ LWRS aperture.  The resulting profile has
an instrumental line spread function significantly broader than usual
for \fuse\ data, and the column density derived for this observation
is an average of the individual sight lines contained within the
aperture (with the absorption profile weighted by the flux
distribution of the targets).  We do not make use of the Milky Way
column densities derived for the NGC~346 sight lines presented in
\citetalias{p2} in this work.  
\label{fig:smcnav}}
\end{figure}

Table \ref{tab:smcovi} gives the Milky Way \ovi\ column densities
derived from an integration of the apparent column density profiles of
the 1032.926 \AA\ transition.  The \ovi\ \nav\ profiles for the SMC
sight lines are shown in Figure \ref{fig:smcnav}.  For the SMC sight
lines, contamination of the \ovi\ absorption profiles by \htwo\ is not
as significant as for the LMC sight lines \citepalias[see][]{p2}.  The
\htwo\ has not been removed before integrating the \nav\ profiles.
The separation between Galactic and SMC \ovi\ absorption is much
cleaner along the SMC sight lines than the LMC sight lines.  For this
reason, the error budget includes the effects of changing the upper
integration limit (last column of Table \ref{tab:smcovi}) by only
$\pm5$ \kms.

\section{Total \ion{O}{6} Column Densities and Degree-Scale Variations}
\label{sec:variations}

The column densities quoted in Tables \ref{tab:lmcovi} and
\ref{tab:smcovi} can be seen to vary significantly from sight line to
sight line.  Towards the LMC the full range of \ovi\ columns spans a
factor of $\sim2.8$ range when integrated over the full Milky Way
velocity limits indicated in the table, or a factor of $\sim4.2$ over
$|v| \le 50$ \kms.  The SMC sight lines show a factor of $\sim4.2$
full range as well.  

\begin{figure}
\epsscale{0.7}
\plotone{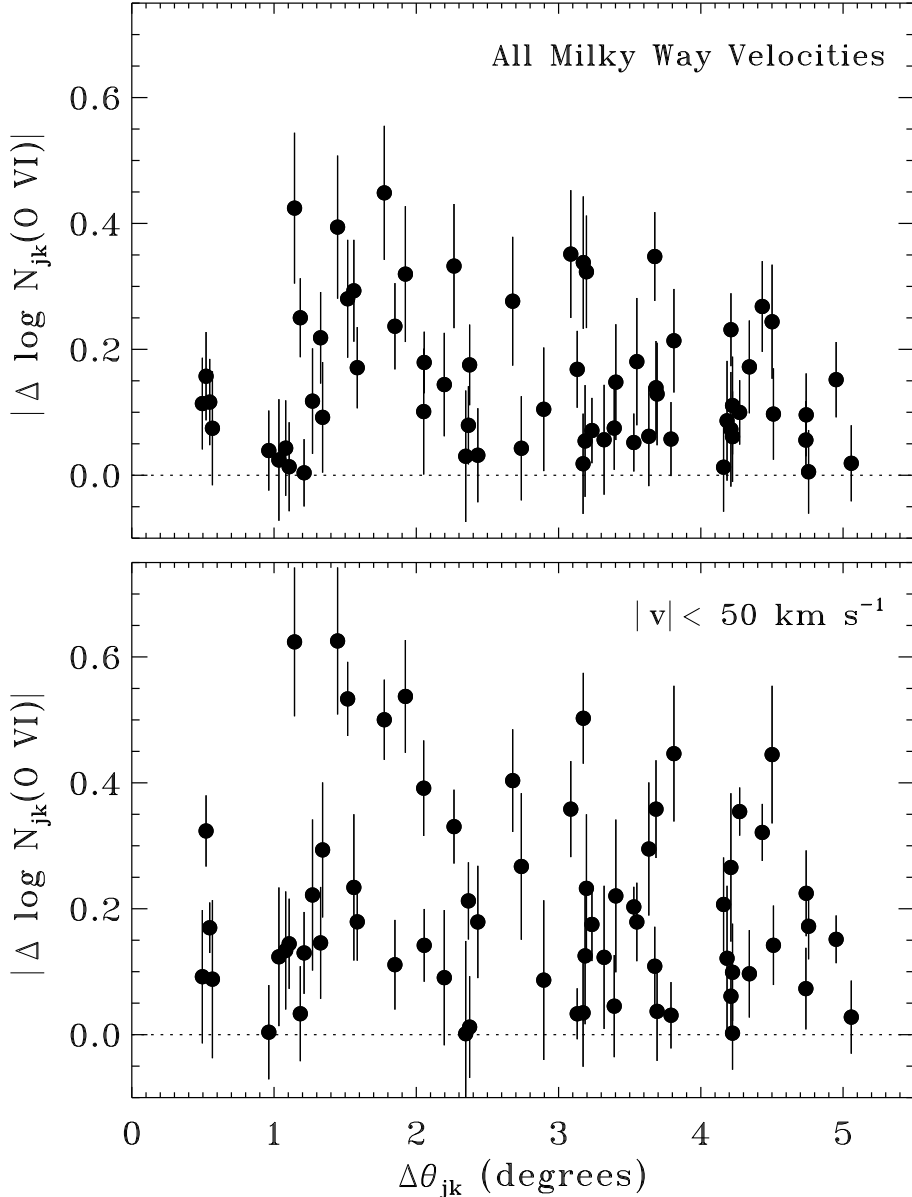}
\caption{The absolute logarithmic difference in \ovi\ column 
densities for each pair of sight lines in Table \ref{tab:lmcovi}.  The
top panel shows the difference in \ovi\ column density integrated over
the full velocity ranges quoted in Table \ref{tab:lmcovi} for each pair
of sight lines.  The bottom panel shows the column density differences
when integrating the apparent column density profiles over the range
$-50 \le v \le +50$ \kms, which excludes much of the expected
intermediate- and all of the expected high-velocity cloud absorption.
\label{fig:lmcvariations}}
\end{figure}

\begin{figure}
\epsscale{0.7}
\plotone{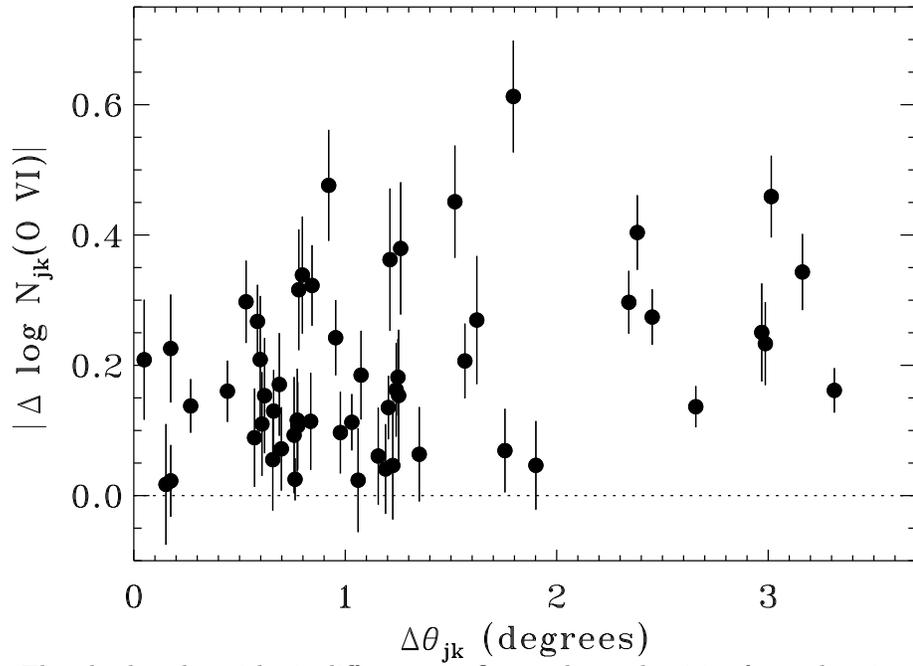}
\caption{The absolute logarithmic difference in \ovi\ column 
densities for each pair of SMC sight lines in Table \ref{tab:smcovi}.
\label{fig:smcvariations}}
\end{figure}

Figure \ref{fig:lmcvariations} shows the absolute value of the
logarithmic differences in column densities for each pairing of sight
lines in our sample, i.e., $| \Delta \log N_{jk}(\mbox{\ovi})| = |
\log N_j(\mbox{\ovi}) - \log N_k(\mbox{\ovi})|$ for each pair of LMC sight
lines $j$ and $k$, versus the angular separation of the pair, $\Delta
\theta_{jk}$.  The top panel of this figure shows the difference in
\ovi\ column density integrated over the full Milky Way velocity range
for each pairing of sight lines, while the bottom panel shows the
column density differences when integrating the apparent column
density profiles over $|v| \le 50$ \kms.  Figure
\ref{fig:smcvariations} shows a similar plot for the SMC sight lines.

Figures \ref{fig:lmcvariations} and \ref{fig:smcvariations}
demonstrate that there are significant column density variations on
all of the angular scales probed by our observations.  The smallest
separations of LMC sight lines are of order $\Delta \theta_{jk} \sim
0\fdg5$, while the smallest angular scales probed by the SMC sight
lines are $\Delta \theta_{jk} \sim 0\fdg05$.  

Table \ref{tab:stats} summarizes the statistics of the \ovi\ column
density measurements towards the LMC and SMC.  For the LMC direction,
the results are shown for integrations over the full Milky Way
velocity range and over $|v| \le 50$ \kms, corresponding to the
analogous entries in Table \ref{tab:lmcovi}.  Because of the presence
of IVCs and HVCs along the LMC sight lines
(\S\ref{subsec:lmcsightline}), the variation in the upper panel of
Figure \ref{fig:lmcvariations} trace variations in both the
low-velocity Milky Way \ovi\ and in the \ovi\ associated with these
anomalous-velocity clouds.  The IVC and HVC absorption along the LMC
sight lines is to a large extent excluded in the integration over the
velocity range $|v| \le 50$ \kms.  The variations over this
low-velocity material are on average larger than those integrated over
the full velocity range (see Table \ref{tab:stats}).  Indeed, it is
interesting to note that the statistics of the LMC and SMC directions
are almost indistinguishable if only the low-velocity gas towards the
LMC is included.

The standard deviations of the individual \ovi\ column density
measurements are a significant fraction of the mean for the LMC and
SMC sight lines.  Table \ref{tab:stats} shows that the standard
deviation of each sample is $27\%$ to $38\%$ of the mean.  For
comparison, the standard deviation of the $N(\mbox{\ovi})$
determinations towards 11 bright AGNs in the study of
\citet{savage00} is $41\%$ of the mean (with $\log \langle N(\mbox{\ovi}) 
\rangle = 14.50$).   Thus, the LMC/SMC measurements that each sample 
sight lines within a few degrees of one another show a degree of
(fractional) variation that is similar to that seen along sight lines
distributed across the entire celestial sphere.  Indeed, compared with
the uncertainties in the individual measurements, the variation
between the LMC and SMC sight lines is more significant than that in
the \citet{savage00} dataset (as judged by the values of $\chi^2$
derived by comparing the data with the mean within each sample).

Figures \ref{fig:lmcvariations} and \ref{fig:smcvariations} show that
significant variations in the \ovi\ column densities of the Galactic
halo ($|\log \Delta N_{jk}(\mbox{\ovi}) | > 0$ at greater than
$2\sigma$ significance) can be found for sight lines over the entire
range of angular scales probed by our observations.  This includes
pairs of sight lines separated by $\la 1^\circ$, many of which which
show greater than 0.1 dex ($\sim25\%$) variations.

\begin{figure}
\epsscale{1.0}
\plotone{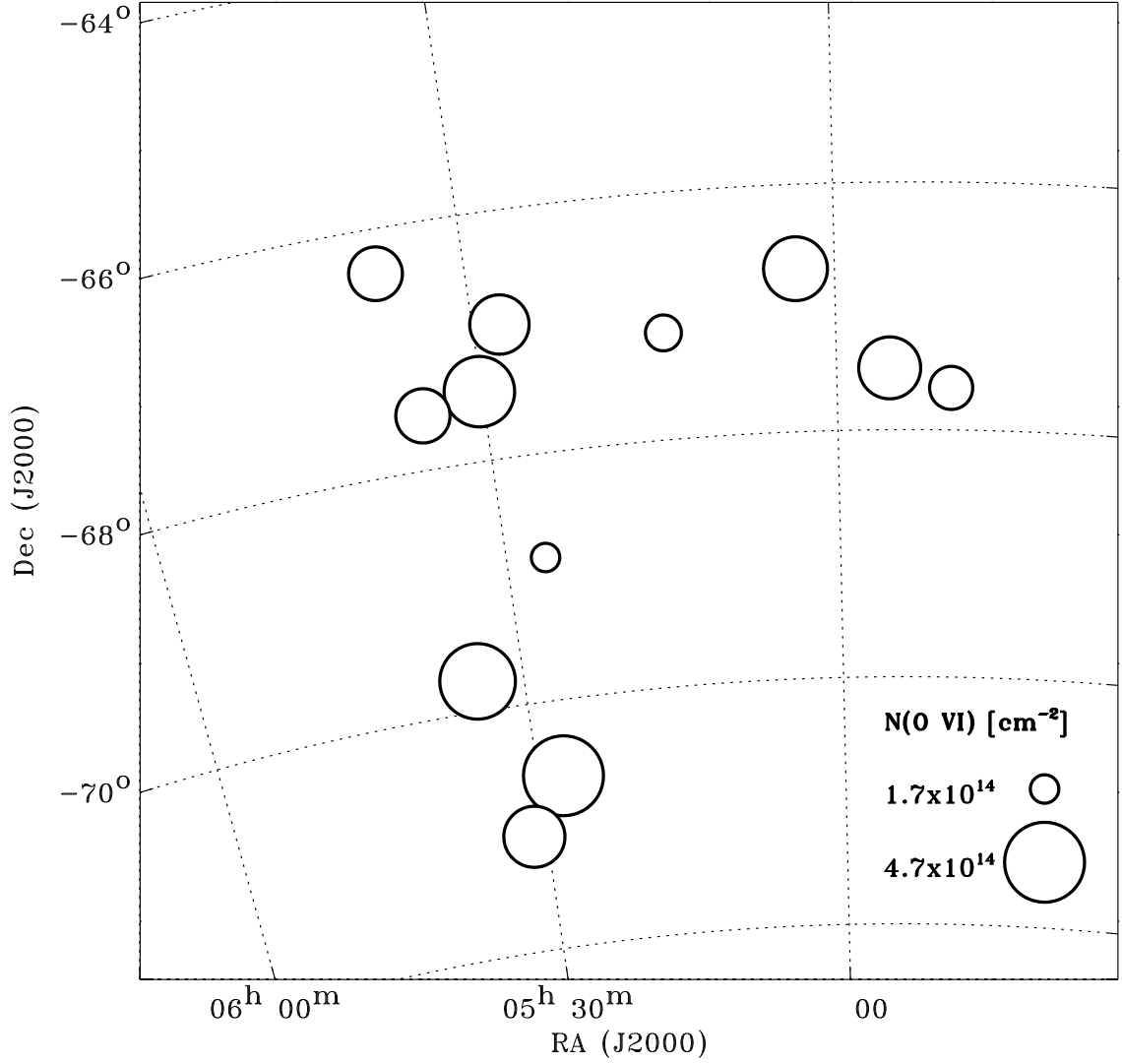}
\caption{A ``map'' of the integrated Galactic \ovi\ column densities
observed towards 12 LMC probes.  Each circle is centered on the
background star and has a radius linearly proportional to the
integrated Milky Way \ovi\ column.  For scale, circles denoting the
minimum and maximum \ovi\ columns are shown in the bottom right.
\label{fig:lmcmap}}
\end{figure}

\begin{figure}
\epsscale{1.0}
\plotone{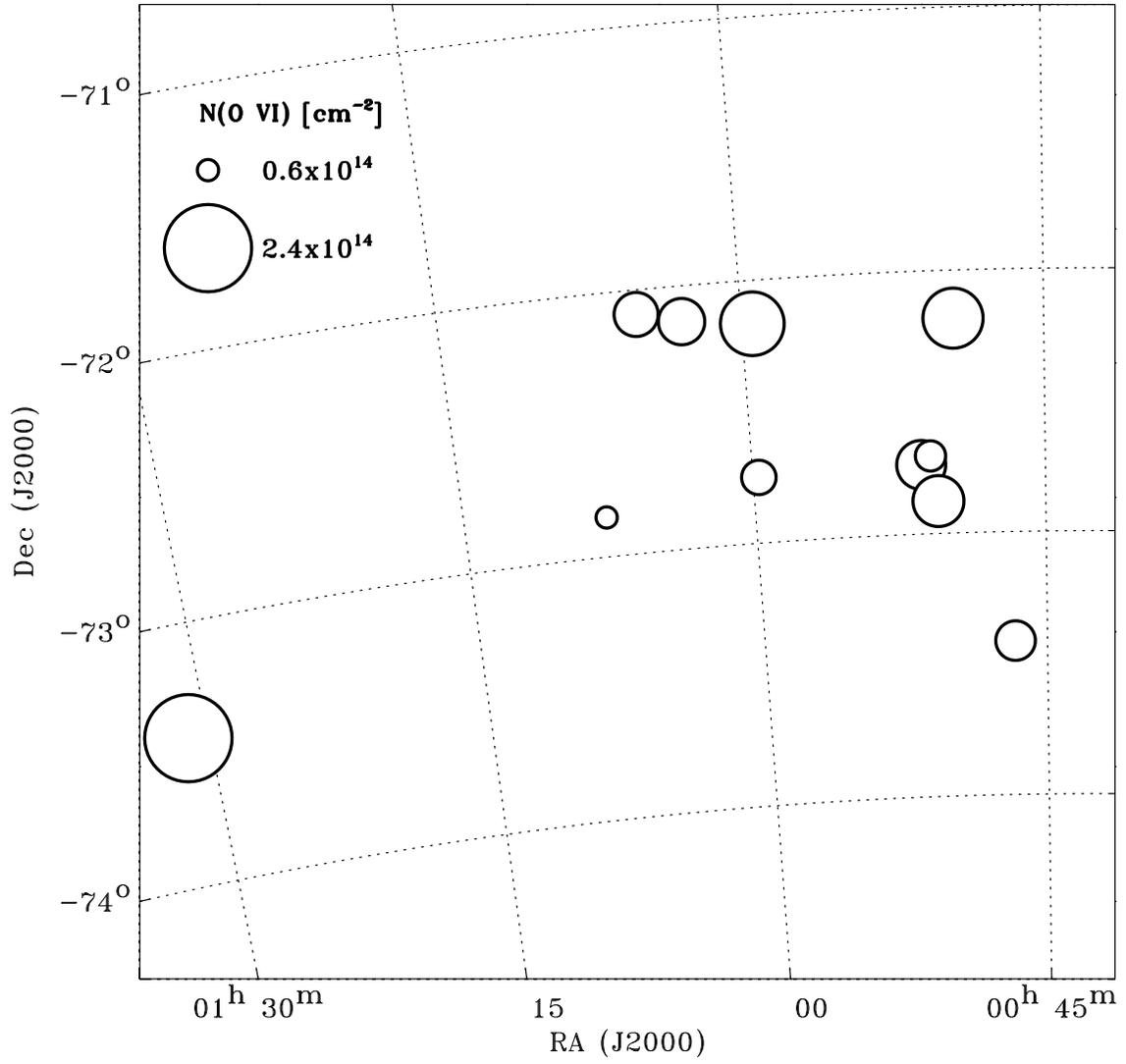}
\caption{As Figure \protect\ref{fig:lmcmap} but for the 11 SMC sight 
lines.
\label{fig:smcmap}}
\end{figure}

To demonstrate the degree of variation of and any systematics in the
distribution of column densities, Figures \ref{fig:lmcmap} and
\ref{fig:smcmap} show ``maps'' of the derived Galactic \ovi\ column
densities towards the LMC and SMC, respectively.  The positions of the
probe stars are marked with circles whose radii are linearly
proportional to the column densities summarized in Tables
\ref{tab:lmcovi} and \ref{tab:smcovi}. Scales showing the minimum and
maximum columns are given.  With only 12 and 11 sight lines, the maps
are too sparsely populated to say whether or not coherent structures
are visible in our data.

\begin{figure}
\epsscale{0.7}
\plotone{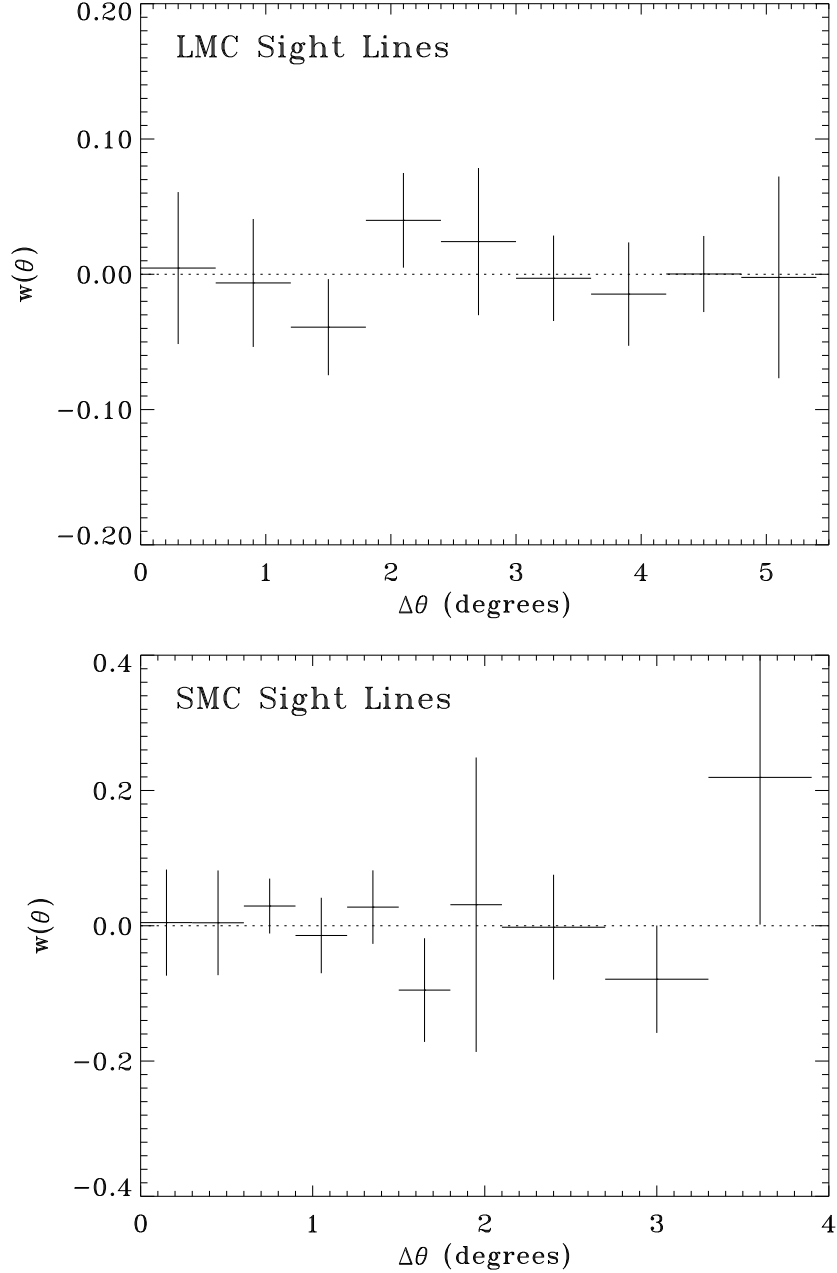}
\caption{The autocorrelation function, $w(\theta)$, for the \ovi\ column 
densities towards the LMC ({\em top}) and SMC ({\em bottom}) derived
as described in the text. The bin spacing for the LMC points is
$0\fdg6$.  The bin spacing for the SMC points is $0\fdg3$ and $0\fdg6$
for $\theta <2^\circ$ and $>2^\circ$, respectively.  The uncertainties
are calculated using simulations (see text) and represent $95\%$
confidence intervals.
\label{fig:acorr}}
\end{figure}

Figure \ref{fig:acorr} shows the angular autocorrelation function
(ACF), $w(\theta)$, for the \ovi\ column densities towards the LMC and
SMC.  The ACF was calculated in a manner analagous to the studies of
the X-ray sky by \citet{chen94} and \citet{kuntz01}.  We define the
ACF as:
\begin{equation}
w(\theta) \equiv 
	\frac{\sum\limits_i \sum\limits_{j\neq i} (N_i-\langle N \rangle)
		(N_j-\langle N \rangle) (\sigma_i \sigma_j)^{1/2}}
	{\langle N \rangle^2 \sum\limits_i \sum\limits_{j\neq i}
		(\sigma_i \sigma_j)^{1/2}}.
\end{equation}
The summation is calculated for each pair of sight lines $i$ and $j$
separated by an angle $\theta$ on the sky. In this expression $N_i$
and $N_j$ are the \ovi\ column densities and $\sigma_i$ and $\sigma_j$
the column density uncertainties for the sight lines, while $\langle N
\rangle$ is the average column density for the entire sample of 12 or 11 
stars.  The $w(\theta)$ data displayed in Figure \ref{fig:acorr} have
been summed over $0\fdg6$ bins for the LMC dataset; the summation for
the SMC dataset is over $0\fdg3$ bins for separations smaller than
$2^\circ$ and $0\fdg6$ bins for larger angular separations.  The
uncertainties in $w(\theta)$ represent $95\%$ confidence limits.
These were determined through 1000 simulations wherein the column
densities were randomly shuffled among the observed directions and
$w(\theta)$ rederived, similar to the method described by
\citet{carrera91}.  The confidence limits shown in Figure
\ref{fig:acorr} include $95\%$ of the simulated $w(\theta)$ values.  This
approach to estimating uncertainties has the advantage that the
simulated data are drawn from the same population as the real data,
and hence no intrinsic column density distribution function need be
assumed.  To remove any systematic effects from the ACF, we have
subtracted the mean simulated $w(\theta)$ from the real values, though
the mean simulated values were always very small compared with the
uncertainties.

Figure \ref{fig:acorr} shows that there is little evidence for
significant power at any scale in the ACF for the \ovi\ column
densities towards the LMC and SMC.  Indeed, those points that seem to
be different from $w(\theta) = 0$ are not robust to changes in the bin
spacing.  When binned over the same angular scale, the LMC and SMC
$w(\theta)$ values are completely consistent over all scales probed.

There is no evidence for significantly greater power in the ACFs on
small angular scales compared with large scales along these two sight
lines.  We caution that the small number of sight lines makes the
uncertainties large over all scales.  Furthermore, the ACFs are
insensitive to structure on scales smaller than the adopted bin size
for each sight line.  Thus, both larger numbers of sight lines and
finer sampling would be needed to probe the smallest structures.
However, our data could be used with further measurements of \ovi\
towards AGNs to constrain the largest-scale \ovi\ structures in the
sky.

\section{Kinematics of Gas in the Galactic Halo}
\label{sec:kinematics}

The kinematic profiles of the \ovi\ absorption lines towards the LMC
and SMC could, in principle, yield information on the distribution of
gas in the Galactic halo assuming it corotates with the underlying
thin disk and adopting an appropriate rotation curve
\citep*[see, e.g.,][]{savagedeboer81,savagesembachlu97}.  Examining
the distribution of apparent column density versus velocity for LMC
and SMC sight lines in this manner can yield estimates of the
turbulent velocity, midplane density, and scale height of the
absorbing layer.

\begin{figure}
\epsscale{0.7}
\plotone{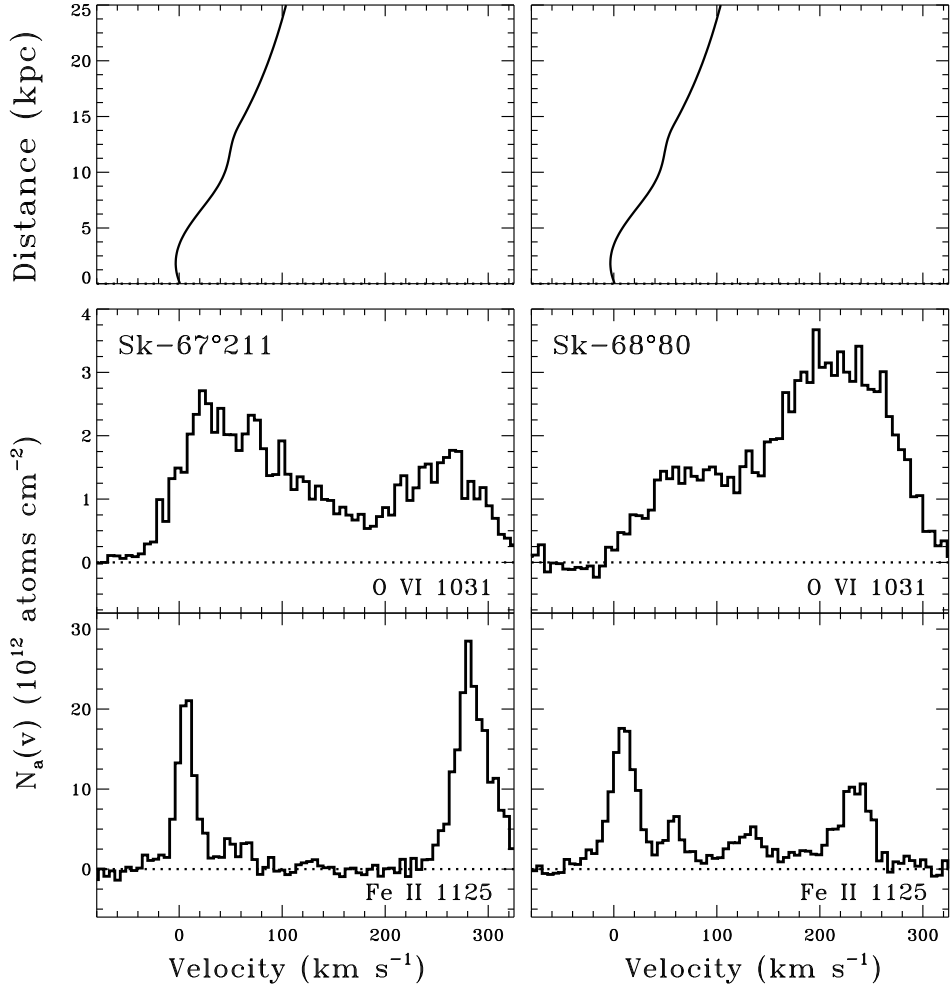}
\caption{Apparent column density profiles for interstellar \ovi\ and 
\feii\ towards the stars \sk67\dg211 and \sk68\dg80 in the LMC (separated 
by $1\fdg5$ on the sky).  Also shown in the top panel is the Galactic
rotation curve as a function of distance from the Sun in the direction
of the LMC from \citep{clemens85}.  Absorption from intermediate- and
high-velocity clouds can be seen in the \feii\ profiles towards both
stars (the weak high-velocity absorption towards \sk67\dg211 is
confirmed in stronger \feii\ transitions).  The apparent column
density profiles are likely lower limits in the cores of the \feii\
lines due to the presence of unresolved saturated structure.
\label{fig:lmcprofiles}}
\end{figure}

There are several caveats to this approach for the particular sight
lines studied here, however.  Figure \ref{fig:lmcprofiles} shows a
comparison of the apparent column density, \nav, profiles of \ovi\ and
\feii\ (derived from the 1125.448 \AA\ transition) for the sight lines 
towards \sk67\dg211 and \sk68\dg80 in the LMC.  The \feii\ absorption
mostly traces gas associated with the neutral ISM (i.e., \HI -bearing
gas).  Also shown is the \citet{clemens85} rotation curve for the
direction of the LMC.\footnote{The reader should be aware that the
velocities presented in the \ovi\ and \feii\ profiles are expected to
be accurate to $\pm10$ to 15 \kms\ when comparing the observed
profiles with the displayed rotation curve.}  These plots demonstrate
the difficulty in analyzing the \ovi\ kinematics in this direction.
Although the sharp \feii\ absorption in the main Milky Way component
suggests the \feii -bearing material is confined to relatively small
distances from the midplane \citep[consistent with the results
of][]{edgarsavage89}, determining the extent of the Milky Way halo
\ovi\ absorption is difficult given the overlap with IVC, HVC, and LMC
material.

Furthermore, examining Figure \ref{fig:lmcprofiles} shows that the \ovi\
\nav\ profiles for these two sight lines are significantly different
at even low velocities, i.e., velocities that should be free from
contamination by IVC and HVC contamination.  The \sk67\dg211 sight
line shows much larger apparent column densities over all velocities
$v \le +50$ \kms\ than the sight line towards \sk68\dg80.
Furthermore, the shape of the profiles is quite different over the
lowest velocities, with the \sk67\dg211 profile exhibiting a sharp
peak at low positive velocities and significant absorption at negative
velocities (neither of which are easily identified along the
\sk68\dg80 sight line).  Given the differences seen between the two
LMC sight lines examined above, which are separated by $1\fdg5$ on the
sky, models of smooth, plane-parallel halo material corotating with
the underlying Galactic disk do not appropriately describe the
complicated nature of the observed \ovi\ profiles.  

\begin{figure}
\epsscale{0.7}
\plotone{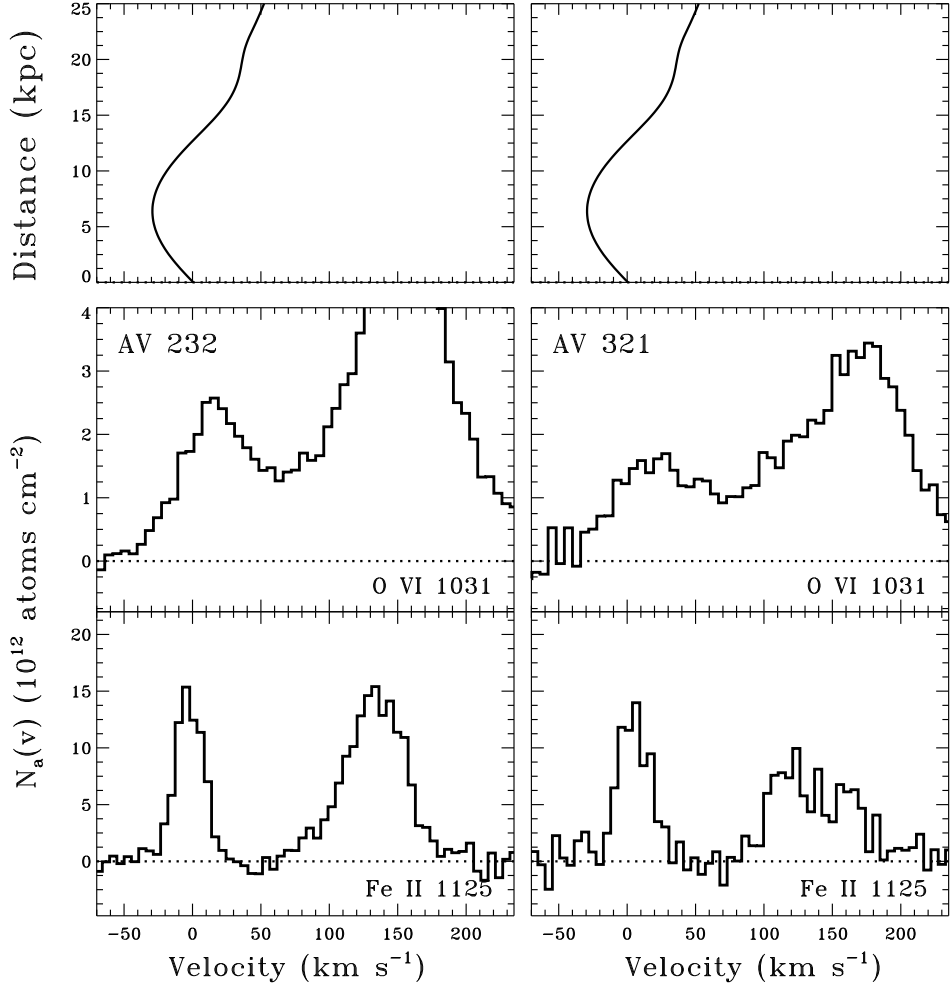}
\caption{Apparent column density profiles for interstellar \ovi\ and 
\feii\ towards the stars AV 232 and AV 321 in the SMC (separated by 
$0\fdg3$ on the sky).  Also shown in the top panel is the Galactic
rotation curve as a function of distance from the Sun in the direction
of the SMC from \citep{clemens85}.  The apparent column density
profiles are likely lower limits in the cores of the \feii\ lines due
to the presence of unresolved saturated structure.
\label{fig:smcprofiles}}
\end{figure}

Figure \ref{fig:smcprofiles} shows the \ovi\ and \feii\ profiles for
the stars AV 232 and AV 321 in the SMC.  These stars are separated by
only $0\fdg3$ on the sky and again significant variations between the
sight lines are seen at low velocity in the \ovi\ \nav\ profiles.  The
two \nav\ profiles exhibit differences not only in the peak and total
\ovi\ column densities, but also in the velocity distribution of the 
gas.  While there is no clear evidence for IVC absorption in these
directions \citep[although][have found weak IVC absorption in the
direction of the SMC star Sk~108]{welty97}, absorption associated with
the SMC begins to overlap the Milky Way absorption at velocities
nearing +100 \kms.  The low-velocity tail of the broad \ovi\
distribution in the SMC likely overlaps the Galactic \ovi\ at
velocities below this. 

While the derivation of reliable estimates of scale height, midplane
densities, and turbulent velocities using the \ovi\ absorption in
these directions may give ambiguous results, examination of the
kinematic profiles of Milky Way \ovi\ seen towards the Magellanic
Clouds do lead to several important conclusions regarding the nature
of highly-ionized Galactic halo gas.  That the individual absorbing
components are seen more distinctly in the \feii\ profiles (Figure
\ref{fig:lmcprofiles}) implies the \ovi\ absorption contains more
and/or broader component complexes than the \feii.  This is consistent
with our expectations given the temperature of the gas the two species
are thought to trace.  

In the direction of the SMC, the Milky Way \ovi\ absorption is clearly
broader than the \feii\ absorption.  This suggests that the \ovi\
arises in a layer much more extended than that traced by \feii, which
has a vertical extension similar to that of \HI\
\citep{edgarsavage89}.  Towards the LMC, the fraction of the gas
arising at velocities associated with IVC and HVC absorption is quite
different for the \ovi\ and \feii\ profiles.  A larger fraction of the
\ovi\ absorption is found at high velocities.  Given that the IVC and
HVC material towards the LMC likely resides far from the Galactic
plane, this implies a greater fraction of the \ovi\ is found in the
Galactic halo than \feii.

This all suggests that while the \ovi -bearing medium is quite
inhomogeneous, the current observations certainly suggest the \ovi\
layer of the Galaxy is significantly different than the \feii -bearing
layer.  \ovi\ is found at preferentially larger distances from the
Galactic plane than is the \feii.  Stated another way, the effective
scale height of the \ovi -bearing gas is larger than that of the \feii
-bearing material.  

\section{Discussion}
\label{sec:discussion}

We have presented measurements of \ovi\ in the Galactic halo observed
towards early-type stars in the LMC and the SMC.  These show
significant \ovi\ column density variations on degree scales (and
smaller) over velocity ranges corresponding to gas in the Galactic
disk and halo.  The dispersion of halo \ovi\ measurements is
$\sim27\%$ to $\sim38\%$ of the averages in these directions.  The
ratio of the highest to lowest \ovi\ column densities ranges from 2.8
to 4.2, with the latter applying to the low-velocity gas towards the
LMC and the entire Milky Way range towards the SMC.  

Several properties of \ovi\ in the Galactic halo can be inferred from
the properties of these sight lines towards the Magellanic Clouds.
The most fundamental inference to be drawn from our observations is
that the \ovi -bearing medium of the Galactic halo is better described
by spatially-patchy distribution than by a smooth layer \citep[e.g.,
like that discussed by][]{spitzer56}.  That the medium is patchy
favors models in which the \ovi\ resides in clouds of $3\times 10^5$ K
material or in the interfaces between cooler ($<10^5$ K) clouds and a
hotter ($\ga 10^6$ K) medium.  The evidence for this conclusion comes
both from the observed column density variations and the kinematic
differences between sight lines (see Figures \ref{fig:lmcnav} and
\ref{fig:smcnav}).

The greater velocity breadth of \ovi\ than \feii\ in the direction of
the SMC and the much larger contribution of intermediate- and
high-velocity halo clouds to the \ovi\ profiles than to the \feii\
profiles towards the LMC imply that the Galactic \ovi\ in these
directions is extended further from the plane than the \feii.
Effectively, these inferences imply the scale height of Galactic \ovi\
is larger than that of \feii.  Although the models of
smoothly-distributed, exponentially-stratified gas layers assumed for
determining the properties of the \ovi\ layer seem inappropriate
(\S\ref{sec:kinematics}), our data do require that the Galactic \ovi\
and \feii\ be distributed in very different ways.

That the Galactic halo \ovi -bearing medium is poorly described by a
smoothly distributed layer of gas is a conclusion one must also draw
from the \citet{savage00} observations of \ovi\ absorption towards
extragalactic FUV sources.  However, the current observations show
that the distribution of \ovi\ within the halo must have significant
variations over much smaller angular, and hence physical, scales.  To
what physical scales might the observed variations correspond?  The
distribution of \ovi\ absorption seems to be more tightly concentrated
towards the Galactic midplane than that of \civ\ and \siiv\
\citep{savage00}.  The latter ions have distributions with best fit
scale heights of $\sim4.4$ and 5.1 kpc \citep{savagesembachlu97} --
under the assumption that an exponentially-stratified layer with some
degree of irregularities (patchiness) is a good description of the
intrinsic distribution.  If we assume that the observed \ovi\
absorption in the Galactic halo towards the Magellanic Clouds resides
at heights lower than $z\sim5$ kpc (i.e., within one \siiv\ scale
height), then the $0\fdg5 - 5\fdg0$ scales probed by our LMC
observations correspond to linear distances of $\ll 80 - 800$ pc while
the $0\fdg05 - 3\fdg3$ scales probed by the SMC measurements
correspond to distances of $\ll 6-400$ pc.  Much of the gas is likely
to be significantly closer to the Sun than this limit, making these
estimated size scales upper limits.  In particular, if the effective
scale height of the \ovi\ distribution is $h_z \sim 2.7$ kpc
\citep[as derived by][]{savage00}, then  $\sim85\%$ of the 
Galactic \ovi\ is expected to be at heights smaller than the adopted 5
kpc.

If we assume that the \ovi\ variations are due to clouds at this
$\sim5$ kpc distance from the plane with angular sizes approximately
equal to the transverse separation of the pairing of stars observed
(giving physical sizes of $80-800$ pc towards the LMC and $5-400$ pc
towards the SMC), we can estimate the required densities of \ovi\ for
the observed column density differences between the sight lines.  The
mean density derived under these assumptions is $\langle n_{\rm O VI}
\rangle \sim 1.5\times10^{-7}$ \percc\ for the LMC sight lines with
significant ($\ge 2\sigma$) column density differences; the full range
of densities is $\sim(0.4 \ {\rm to } \ 4.4) \times10^{-7}$ \percc.
Towards the SMC the mean density derived in this manner is $\langle
n_{\rm O VI} \rangle \sim 2.7\times10^{-7}$ \percc\ with a full range
of $\sim(0.6 \ {\rm to } \ 27) \times10^{-7}$ \percc.  These values
give estimated particle densities of $\langle n_{\rm H}
\rangle \sim 1.4\times10^{-3}$ \percc\ and $\sim2.5\times10^{-3}$ \percc\ 
for the LMC and SMC directions, respectively, assuming the abundance
of oxygen relative to hydrogen is $5.5\times10^{-4}$
\citep{holweger01} and the fraction of oxygen in the form of O$^{+5}$
is $\la 0.2$ \citep{sutherlanddopita93}.  

The average density estimates can be used to estimate the pressure of
the \ovi -bearing gas.  For $P/k = nT = 2.3 n_{\rm H} T$, where the
factor of 2.3 allows for the total number of particles in a fully
ionized gas containing $10\%$ helium, we obtain $\langle P/k \rangle
\sim 10^3$ to $1.7\times 10^3$ K \percc\ for the directions to the LMC
and SMC, respectively.  These estimates adopt the average densities
derived in the previous paragraph, and we have assumed $T =
3\times10^5$ K, the temperature where \ovi\ peaks in abundance in
collisional ionization equilibrium.  The densities and pressures
derived in this way will be higher for clouds closer than the assumed
5 kpc height from the plane of the Galaxy.

This approach to estimating densities and pressures is overly
simplistic, and we caution the reader not to put too much emphasis on
these idealized calculations -- neither the true distances to nor the
morphology of the absorbing material is known.  If the morphology of
the \ovi -bearing ISM is more sheet- or filament-like, then much of
the variation in column density will be the result of the projection
of the individual absorbing components along the line of sight
\citep[see, e.g.,][]{heiles97}.  Indeed, the interpretation of the 
absorption arising in individual clouds at a common distance is highly
unsatisfactory given the apparent lack of any any preferred scales for
the observed variability.

We suggest that the observations presented in this work, which show
degree-scale column density and kinematic variations in the \ovi
-bearing medium of the Galactic halo, fit readily into pictures of the
Galactic halo wherein the \ovi\ is contained in complicated cloud- or
sheet-like distributions of material.  In particular, models that
place the \ovi\ gas in interfaces between more neutral clouds or
sheets and a more pervasive very hot ($\ga10^6$ K) ISM would seem to
allow for the observed variations so long as there are not so many
such interfaces along typical paths through the halo that the expected
variations along random adjacent sight lines is expected to be small.

The morphological information contained in the absorption line
datasets presented here is incomplete, probing only random pencil
beams through the highly-ionized halo.  With our sparsely-populated
``maps'' it is difficult to discern the structure of the halo \ovi.
Our observations do give the first hint of the complexity of the
small-scale distribution of hot, highly-ionized material within the
halo.  The structures giving rise to the \ovi\ column density
variations demonstrated in this work will be accessible to future
\ovi\ emission-line imaging programs.  Indeed, that we find 
significant variations on the smallest angular scales probed suggests
there is a wealth of structure present in the highly-ionized sky;
imaging data may reveal how this highly-ionized material is related to
other phases of the ISM.

We also note that the degree of structure demonstrated by our \ovi\
absorption measurements has important implications for understanding
current observations of diffuse \ovi\ emission.  The comparison of
\ovi\ emission line intensities with the \ovi\ column density along
the same line of sight can yield estimates of the physical properties
of the \ovi -bearing gas, e.g., the density and pressure
\citep{shelton01,dixon01}.  However, the determination of 
the appropriate column density for comparison with an emission line
measurement is difficult.  Our observations (e.g., Figures
\ref{fig:lmcvariations} and
\ref{fig:smcvariations}) suggest that \ovi\ absorption line 
measurements along sight lines even a few arcminutes from the
direction of an emission line measurement can only be assumed to yield
column densities within a factor of approximately two of the value
appriopriate for the emitting region.  The large degree of variation
in the absorption line column densities along closely-spaced sight
lines through the Galactic halo must be considered when assessing the
systematic uncertainties in comparing emission and absorption line
measurements of \ovi.  However, one may look at this in another way:
our analysis suggests that sight lines several degrees apart give
column densities that are expected, on average, to give results that
are no less consistent than sight lines separated by less than a
degree.  That is, unless the opportunity exists to observe \ovi\
emission and absorption along essentially the same sight line, the
determination of an \ovi\ column density several degrees away from an
emission line measurement will provide no worse an indicator than an
\ovi\ column density determined for a sight line tens of arcminutes away.

\section{Summary}
\label{sec:summary}

We have presented observations of Galactic halo \ovi\ absorption
towards 12 and 11 early type stars in the Large and Small Magellanic
Clouds, respectively.  These observations reveal strong column density
variations on degree scales.  The principle results of this study are
as follows.

\begin{enumerate}

\item The Galactic halo \ovi\ probed by our observations shows significant 
structure on all of the scales probed by our observations
($0\fdg05-5\fdg0$).  There seems to be no preferred angular scale for
\ovi\ variations over this range.  The transverse size-scales 
probed by our observations are likely $\ll 6$ to 800 pc.

\item The Galactic \ovi\ in the directions of the Magellanic Clouds is 
very patchy, with the standard deviations of the measurements being
equivalent to $\sim26\%$ to $38\%$ of the average values in these
directions.  The maximum variations correspond to factors of $\sim2.9$
to 4.2.

\item Galactic \ovi\ is distributed in a very different manner than \feii.
The \ovi -bearing layer of the Galaxy is significantly more extended
than the \feii -bearing layer, which has a vertical extent similar to
that of \HI.  However, the existence of large variations in the \ovi\
column densities over very small angular scales implies that models of
a smoothly-varying, exponentially-stratified halo are poor
descriptions of the true distribution of highly-ionized gas.  The
meaning of detailed quantities derived by comparing such models with
the observed \ovi\ kinematics along a given sight line (or sight
lines) is ambiguous at best.

\item The existence of large column density variations over small angular
scales can be used to constrain the origin of Galactic halo \ovi\
ions.  Models in which the \ovi -bearing medium is composed of
complicated cloud- or sheet-like distributions of material
(particularly narrow interfaces) should be considered more likely than
smoothly-distributed layers of highly-ionized material in light of our
observations.

\end{enumerate}

\acknowledgements

This work is based on data obtained for the Guaranteed Time Team by
the NASA-CNES-CSA FUSE mission operated by the Johns Hopkins
University. Financial support to U. S. participants has been provided
by NASA contract NAS5-32985.  JCH and KRS recognize support from NASA
Long Term Space Astrophysics grant NAG5-3485 through the Johns Hopkins
University.

%%%%%%%%%%%%%%%%%%%%%%%%%%%% Begin Tables  %%%%%%%%%%%%%%%%%%%%%%%%%%%%

\begin{deluxetable}{llllllll}
\tabletypesize{\scriptsize}
\tablenum{1}
\tablecolumns{7}
\tablewidth{0pt}
\tablecaption{Interstellar \ion{O}{6} in the Halo of the 
		Milky Way Towards the LMC \label{tab:lmcovi}}
\tablehead{
\colhead{Star} & 
\colhead{$\alpha$} & \colhead{$\delta$} &
\colhead{$(l,b)$} &
\colhead{$W_\lambda$\tablenotemark{a}} &
\multicolumn{2}{c}{$\log N(\mbox{\ion{O}{6}})$\tablenotemark{b}} &
\colhead{$v_-,v_+$\tablenotemark{c}} \\
\cline{6-7}
\colhead{} & 
\colhead{(J2000)} & \colhead{(J2000)} & 
\colhead{[deg.]} & 
\colhead{[m\AA]} &
\colhead{$[-50,+50]$} & \colhead{$[v_-,v_+]$} &
\colhead{}
}
\startdata
 Sk--67\dg05\tablenotemark{d} & $04^h\, 50^m\, 18\fs8$ & $-67^\circ\, 39\arcmin \, 38\farcs2$ & $278.89, -36.32$ &  $ 249\pm10$ & $13.79\pm0.03$ & $14.40\pm0.05$ & $ -50,+180$  \\
 Sk--67\dg20 &  $04$  55  31.5 & $-67$ 30 00.9 & $278.53, -35.89$ &  $ 324\pm16$ & $14.11\pm0.05$ & $14.56\pm0.05$ & $ -35,+175$  \\
 Sk--66\dg51 &  $05$  03  10.1 & $-66$ 40 53.9 & $277.32, -35.38$ &  $ 317\pm21$ & $13.97\pm0.05$ & $14.57\pm0.05$ & $ -25,+180$  \\
 Sk--67\dg69 &  $05$  14  20.1 & $-67$ 08 03.5 & $277.57, -34.22$ &  $ 200\pm15$ & $14.00\pm0.05$ & $14.32\pm0.04$ & $ -40,+160$  \\
 Sk--68\dg80 &  $05$  26  30.4 & $-68$ 50 26.6 & $279.34, -32.79$ &  $ 164\pm 9$ & $13.61\pm0.05$ & $14.22\pm0.09$ & $ -15,+140$  \\
 Sk--70\dg91 &  $05$  27  33.7 & $-70$ 36 48.3 & $281.40, -32.40$ &  $ 377\pm11$ & $14.11\pm0.03$ & $14.67\pm0.06$ & $ -40,+175$  \\
Sk--66\dg100 &  $05$  27  45.5 & $-66$ 55 14.9 & $277.06, -32.96$ &  $ 308\pm32$ & $14.15\pm0.07$ & $14.54\pm0.06$ & $ -35,+175$  \\
Sk--67\dg144 &  $05$  30  12.2 & $-67$ 26 08.4 & $277.63, -32.66$ &  $ 333\pm29$ & $14.23\pm0.10$ & $14.61\pm0.07$ & $ -40,+175$  \\
 Sk--71\dg45 &  $05$  31  15.5 & $-71$ 04 08.8 & $281.87, -32.02$ &  $ 322\pm 7$ & $13.94\pm0.02$ & $14.55\pm0.04$ & $ -30,+185$  \\
Sk--69\dg191 &  $05$  34  19.3 & $-69$ 45 10.0 & $280.29, -31.97$ &  $ 352\pm38$ & $14.23\pm0.10$ & $14.64\pm0.08$ & $ -30,+175$  \\
Sk--67\dg211 &  $05$  35  13.9 & $-67$ 33 27.0 & $277.70, -32.16$ &  $ 294\pm 8$ & $14.14\pm0.02$ & $14.50\pm0.02$ & $ -50,+185$  \\
Sk--66\dg172 &  $05$  37  05.5 & $-66$ 21 35.7 & $276.27, -32.10$ &  $ 283\pm21$ & $14.01\pm0.06$ & $14.50\pm0.05$ & $ -45,+195$  \\
\enddata
\tablenotetext{a}{Equivalent widths for the Milky Way material 
	along the observed sightlines with $1\sigma$ error estimates.}
	%The error estimates include the effects of shifting the upper
	%velocity integration limit by $\pm20$ km s$^{-1}$.}
\tablenotetext{b}{\ion{O}{6} column densities for Milky Way material 
	along the observed sight lines with $1\sigma$ error estimates.
	Two column densities are given for each sight line: one
	integrated over the velocity range $v = -50$ to +50 km
	s$^{-1}$, the other integrated over a range $v_-$ to $v_+$,
	where the values of the limits are given in the last column of
	this table.  The errors in the latter case include the effects
	of varying the upper velocity limit by $\pm20$ km s$^{-1}$
	(see text).  In all cases the column densities have been
	derived using observations of the 1031.926 \AA\ transition
	assuming no unresolved saturation is present.  We assume an
	$f$-value of $f = 0.1325$ from the theoretical calculations of
	Yan, Tambasco, \& Drake 1998.}
\tablenotetext{c}{Velocity range over which the full Milky Way  profile 
	was integrated.}
\tablenotetext{d}{The \ion{O}{6} column densities towards Sk--67\dg05
	are taken from Table 2 of Friedman et al. 2000 assuming their
	``upper'' continuum placement.  The ``lower'' continuum
	placement implies a column density 0.08 dex lower than that
	quoted.}
\end{deluxetable}
 %\ref{tab:lmcovi}
\begin{deluxetable}{llllllll}
\tablenum{2}
\tablecolumns{7}
\tablewidth{0pt}
\tablecaption{Interstellar \ion{O}{6} in the Halo of the 
	Milky Way Towards the SMC \label{tab:smcovi}}
\tablehead{
\colhead{Star} & 
\colhead{$\alpha$} & \colhead{$\delta$} &
\colhead{$(l,b)$} &
\colhead{$W_\lambda$\tablenotemark{a}} &
\multicolumn{1}{c}{$\log N(\mbox{\ion{O}{6}})$\tablenotemark{b}} &
\colhead{$v_-,v_+$\tablenotemark{c}} \\
\cline{6-6}
\colhead{} & 
\colhead{(J2000)} & \colhead{(J2000)} & 
\colhead{[deg.]} &
\colhead{[m\AA]} &
\colhead{$[v_-,v_+]$} &
\colhead{}
}
\startdata
       AV 15 &  $0           0^h\, 46^m\, 42\fs1$ & $-73^\circ\, 24\arcmin \, 54\farcs7$ & $303.40, -43.71$ &  $ 109\pm11$ & $14.04\pm0.05$ & $ -45,+ 65$  \\
       AV 75 &  $00$  50  32.5 & $-72$ 52 36.2 & $303.03, -44.25$ &  $ 133\pm14$ & $14.15\pm0.06$ & $ -40,+ 50$  \\
       AV 83 &  $00$  50  52.0 & $-72$ 42 14.5 & $302.99, -44.42$ &  $  84\pm10$ & $13.93\pm0.06$ & $ -50,+ 40$  \\
       AV 95 &  $00$  51  21.5 & $-72$ 44 12.8 & $302.94, -44.39$ &  $ 129\pm18$ & $14.14\pm0.07$ & $ -45,+ 60$  \\
      AV 229 &  $00$  59  27.7 & $-72$ 09 55.0 & $302.07, -44.95$ &  $ 164\pm 4$ & $14.25\pm0.02$ & $ -45,+ 70$  \\
      AV 232 &  $00$  49  30.0 & $-72$ 11 00.0 & $303.14, -44.94$ &  $ 155\pm 5$ & $14.23\pm0.02$ & $ -60,+ 65$  \\
      AV 235 &  $00$  59  42.0 & $-72$ 45 00.0 & $302.08, -44.36$ &  $ 100\pm10$ & $13.98\pm0.05$ & $ -40,+ 75$  \\
      AV 321 &  $01$  02  57.0 & $-72$ 08 09.3 & $301.69, -44.96$ &  $ 130\pm 7$ & $14.11\pm0.04$ & $ -40,+ 75$  \\
      AV 378 &  $01$  05  09.4 & $-72$ 05 35.0 & $301.44, -45.00$ &  $ 123\pm10$ & $14.09\pm0.04$ & $ -40,+ 70$  \\
      AV 423 &  $01$  07  40.4 & $-72$ 50 59.5 & $301.26, -44.23$ &  $  61\pm10$ & $13.77\pm0.08$ & $ -25,+ 40$  \\
      Sk 188 &  $01$  31  06.0 & $-73$ 26 00.0 & $299.06, -43.40$ &  $ 215\pm 8$ & $14.39\pm0.02$ & $ -65,+ 85$  \\
\enddata
\tablenotetext{a}{Equivalent widths for the Milky Way material 
	along the observed sightlines with $1\sigma$ error estimates.}
	%The error estimates include the effects of shifting the upper
	%velocity integration limit by $\pm20$ km s$^{-1}$.}
\tablenotetext{b}{\ion{O}{6} column densities for Milky Way material 
	along the observed sight lines with $1\sigma$ error estimates.
	The column densities have been derived using observations of
	the 1031.926 \AA\ transition assuming no unresolved saturation
	is present and integrating over the range $v_-$ to $v_+$,
	where the values of the limits are given in the last column of
	this table.  The errors include the effects of varying the
	upper velocity limit by $\pm5$ km s$^{-1}$ (see text).  We
	assume an $f$-value of $f = 0.1325$ from the theoretical
	calculations of Yan, Tambasco, \& Drake 1998.}
\tablenotetext{c}{Velocity range over which the full Milky Way  profile 
	was integrated.}
\end{deluxetable}
 %\ref{tab:smcovi}
\begin{deluxetable}{ccccc}
\tablenum{3}
\tablecolumns{5}
\tablewidth{0pt}
\tablecaption{Statistical Properties of Galactic  Halo \ovi\ Towards the LMC and SMC 
	\label{tab:stats}}
\tablehead{
\colhead{} & \multicolumn{2}{c}{LMC} & 
\colhead{} & \colhead{SMC}\\
\cline{2-3} \cline{5-5} 
\colhead{Property} &
\colhead{$[-50, +50]$\tablenotemark{a}} &
\colhead{$[v_-,v_+]$\tablenotemark{b}} &
\colhead{} & \colhead{$[v_-,v_+]$} 
}
\startdata
$\log \langle N(\mbox{\ion{O}{6}}) \rangle$ &
	14.06  & 14.52  &  & 14.13 \\
$\log \langle N(\mbox{\ion{O}{6}}) \sin |b| \rangle$ &
	13.80 & 14.26  &  & 13.97 \\
$\sigma_{N({\rm O \, VI})} / 
	\langle N(\mbox{\ion{O}{6}}) \rangle$\tablenotemark{c} &
	 $36\%$ & $27\%$  & & $38\%$ \\
%$\chi^2$\tablenotemark{d} &
%	
$\langle | \Delta \log  N(\mbox{\ion{O}{6}}) | \rangle$\tablenotemark{d} &
	 0.21 & 0.15 & &    0.20 \\ 
Max. $| \Delta \log  N(\mbox{\ion{O}{6}}) | $ & 
	0.62 & 0.45  &  & 0.62 \\
\enddata
\tablenotetext{a}{This column describes the statistics of the Galactic 
	halo \ion{O}{6} column densities in the direction of the LMC
	when integrated over velocities $|v| \le 50$ km s$^{-1}$.}
\tablenotetext{b}{This column describes the statistics of the Galactic 
	halo \ion{O}{6} column densities in the direction of the LMC
	when integrated over the full range $v_-, v_+$ given in Table
	1.}
\tablenotetext{c}{The standard deviation of the measurements compared 
	with the mean (given in percentage).}
%\tablenotetext{d}{The $\chi^2$ of the distribution of data points 
%	compared with the mean.}
\tablenotetext{d}{The average value of $| \Delta \log 
	N(\mbox{\ion{O}{6}})_{jk} | \equiv | \log N_j(\mbox{\ovi}) -
	\log N_k(\mbox{\ovi})|$ for all pairs of sight lines $j$ and
	$k$.}
\end{deluxetable}
 %\ref{tab:stats}

%%%%%%%%%%%%%%%%%%%%%%%%%%%% Begin Figures %%%%%%%%%%%%%%%%%%%%%%%%%%%%

\end{document}